\documentstyle[aps,prd]{revtex}
\begin{document}
\draft

\preprint{KIAS-P98008}

\title{Background geometry of DLCQ $M$ theory on a $p$-torus 
and holography}
\author{Seungjoon Hyun\footnote{hyun@kiasph.kaist.ac.kr}
    and Youngjai Kiem\footnote{ykiem@kiasph.kaist.ac.kr} }
\address{School of Physics, KIAS, Seoul 130-012, Korea}
\maketitle

\begin{abstract}
Via supergravity, we argue that the infinite Lorentz boost 
along the $M$ theory circle {\em a la} Seiberg toward the 
DLCQ $M$ theory compactified on a $p$-torus ($p<5$) implies 
the holographic description of the microscopic theory.  This 
argument lets us identify the background geometries of DLCQ $M$ 
theory on a $p$-torus; for $p = 0$ ($p=1$), the background 
geometry turns out to be eleven-dimensional (ten-dimensional) 
flat Minkowski space-time, respectively.  Holography for these 
cases results from the localization of the light-cone momentum.  
For $p = 2,3,4$, the background geometries are the tensor 
products of an Anti de Sitter space and a sphere, which, 
according to the AdS/CFT correspondence, have the holographic 
conformal field theory description.  These holographic 
descriptions are compatible to the microscopic theory of 
Seiberg based on $\tilde{M}$ theory on a spatial circle with
the rescaled Planck length, giving an understanding of the 
validity of the AdS/CFT correspondence.  
\end{abstract}



\section{Introduction}

Matrix theory is the leading candidate for the quantum 
formulation of $M$ theory \cite{bfss} \cite{dvv} \cite{matrix}. 
In the eleven dimensional setup, Banks, Fischler,
Shenker and Susskind (BFSS) proposed that the Yang-Mills quantum 
mechanics describes the quantum $M$ theory \cite{bfss}.  Dijkgraaf,
Verlinde and Verlinde (DVV) then formulated the matrix string theory
by considering $M$ theory compactified on a circle, which turned
out to give a second quantized version of string theory \cite{dvv}.  
Shortly thereafter, matrix formulations on a higher torus followed
\cite{wat} \cite{roz} \cite{brs} \cite{ns}.  Susskind noted that 
the light-cone compactification, the discrete
light-cone quantization (DLCQ), of the $M$ theory circle gives 
a formulation for a finite value of $M$-momentum that becomes 
identical to the original matrix theory in the large momentum 
limit \cite{susskind}.  Recently, Seiberg \cite{seiberg}
and Sen \cite{sen} provided us with an understanding
on why the matrix theory formulation works by resorting to the
infinite Lorentz boost along the $M$ theory circle, which
gave us a unified derivation of the matrix theories, and 
clarified the issues that arise when we try to compactify
$M$ theory on a $p$-torus with $p$ larger than three.   

The concept of holography originates from the investigation
of the quantum black hole physics where one supposes that the 
quantum degrees of freedom of a black hole are encoded in 
the black hole horizon \cite{holo}.  The key observation is 
that the black hole horizon plays the role of the infinite 
red-shift surface from the point of view of an outside observer; 
it provides us with the right kinematic restriction to high 
frequencies, and in this situation the four dimensional content 
of the theory may be encoded in the two dimensional submanifold that 
corresponds to a black hole horizon.  In the context of the matrix 
theory, again under the right choice of the kinematic regime, 
corresponding to the infinite momentum frame, the eleven 
dimensional content of the $M$ theory are believed to be 
captured in the transversal nine-dimensional submanifold \cite{bfss}.  
A recent addition to this list of holography is the 
correspondence between the Anti de Sitter (AdS) space and
conformal field theory (CFT) \cite{malda} \cite{witten} 
\cite{polyakov}.  Here under the restriction of 
large $N$, the conformal field theory living on the boundary 
submanifold of an AdS space is conjectured
to be equivalent to the bulk supergravity (or string) theory
on the AdS space \cite{hyun1}.  
From the string theory point of view, a large wrapping number
of $p$-branes and a large momentum number can be interchanged
under an appropriate $U$-duality transformation, suggesting that
the relationship among these holographies may exist.

The convergence of these two lines of investigation is the main theme
of this paper.  Following the prescription of Seiberg \cite{seiberg}
at the level of supergravities, we identify the background 
geometries of DLCQ $M$ theory compactified on a $p$-torus.  
In this process, we get an argument that the Seiberg's DLCQ 
prescription amounts to the realization of the microscopic 
theory {\em via holography}.  Our results, that we derive in detail 
in the following section, are summarized by the table 1. 
\\
\begin{center}
\begin{tabular}{|c|c|c|c|c|} \hline
Torus  & Partons & Background & Holography & Microscopics   \\ 
       &         & geometry   &            &                \\ \hline 
$T^0$  & $D$-particle $\rightarrow$ $M$-momentum & $R^{1,10}$ 
       & LCML  & BFSS $R^{9N}/ S_N$      \\ \hline
$T^1$  & $F$-string momentum ($D$-string) & $R^{1,9}$ 
       & LCML & DVV $R^{8N}/ S_N$  \\ \hline
$T^2$  & $D$-membrane $\rightarrow$ $M$-membrane & $AdS_4 \times S^7$ 
       & $AdS_4/CFT_3$  &  $D=3$ SYM (Infrared limit) \\ \hline
$T^3$  & $D$-threebrane  & $AdS_5 \times S^5$ 
       & $AdS_5/CFT_4$  & $D=4$ SYM (Infrared limit) \\ \hline
$T^4$  & $D$-fourbrane $\rightarrow$ $M$-fivebrane & $AdS_7 \times S^4$ 
       & $AdS_7/CFT_6$  & $D=6$ (2,0) theory (Infrared limit) \\ \hline
\end{tabular}
\\
[.3cm]
 Table 1.  The light-cone momentum localization is denoted as
LCML. 
\end{center}
For $T^p$ with even $p$, the background geometry is 
eleven-dimensional, corresponding to type IIA/$M$ theory side 
description, whereas for odd $p$, it becomes ten-dimensional 
corresponding to the possibility of type IIB theory side description.
For $p=0,1$, we demonstrate how the holography is realized at
the level of the background geometry.  The key feature here is
that, even when we have a microscopic description of the theory in the
infinite momentum frame, the background geometry, Minkowski space-time
as it turns out, retains the full covariance symmetry $SO(1,10)$ 
($SO(1,9)$) and full supersymmetry, as well as being classically 
singularity-free.  Seiberg's argument actually
enables us to identify the underlying microscopic hamiltonian
in these cases; especially for $p=1$, the justification for the
DVV hamiltonian is given along this line.  For $p=2,3,4$, as was 
first noticed in \cite{hyun} by one of us (SH), we find that
the background geometry is an AdS space tensored with
a sphere.  Again, Seiberg's argument lets us identify the microscopic
description of the theory and, from our point of view, this
consideration yields a heuristic justification of 
the AdS$_{(4,5,7)}$/CFT$_{(3,4,6)}$ correspondence in the 
decompactification limit \cite{hyun}.  Similar to the case of $p=0,1$, 
the non-asymptotically flat background geometries in these cases 
retain the full supersymmetry and they have 
an appropriate space-time covariance, as well as being
classically singularity-free.  In the last section, we discuss 
implications of our analysis, comment on the recent observation
by Vafa that AdS$_3$/CFT$_2$ correspondence appears 
incomplete \cite{vafa}, and suggest further lines of development. 

\section{Background geometry of DLCQ $M$ theory on $T^p$}

We start by setting up the notations reviewing 
the $D$-particle solutions on a $p$-torus \cite{hstro}
and the $T$ duality on
them to get $D$ $p$-brane solutions.  Then we consider the infinite 
Lorentz boost along the $M$ theory circle {\em a la} Seiberg to 
formulate the DLCQ $M$ theory.  The analysis for the $p$-torus 
compactification for $p=0,1,2,3,4$ will follow. 

\subsection{Review : $D$-particles on a small $T^p$ and $T$ duality}
An efficient method of calculating the space-time geometry
produced by $D$-particles compactified on $T^p$ is to start
from the description of $D$-particles in a non-compact
space-time.  The metric for $D$-particles in type IIA supergravity 
in a ten-dimensional non-compact space-time is 
\begin{equation}
ds^2_{10} = - \frac{1}{\sqrt{f_0}} dt^2 + \sqrt{f_0} 
    (dx_1^2 + \  \cdots \  + dx_9^2  ),
\label{d0metric}
\end{equation}
where $f_0$ is a nine-dimensional harmonic function
\begin{equation}
 f_0 = 1 + g_s l_s^7
\sum_{n=1}^{N} \frac{Q_n}{| \vec{x} - \vec{x}_n |^7}  .
\label{f0} 
\end{equation}
Here $\vec{x}$ denotes a nine-dimensional spatial point, and
$\vec{x}_n$ and $Q_n$ represent the position of $D$-particles and
the coincident $D$-particle number at $\vec{x}_n$, 
respectively. Due to the supersymmetry of this configuration,
we can simply linearly-superpose the contribution to $f_0$
from each $D$-particle.  The dilaton $\phi$ and 
the R-R one-form gauge field $A$ are given by 
\begin{equation}
 e^{- 2 \phi} = g_s^{-2} f_0^{-3/2} \ \ , \ \
   A_t = 1 - \frac{1}{f_0} , 
\label{rr}
\end{equation}
where the string coupling constant $g_s$ satisfies 
$g_s = e^{\phi_{\infty}}$.  The parameter $l_s$ denotes the
string scale.  

We now suppose that the coordinates 
$x_{10-p}, \ \cdots \ $, and $x_9$ are 
compactified via the identification $x_i \simeq x_i + R_i$, where
$R_i$ denotes the circumference of each circle of $p$-torus $T^p$.
This implies that we have to include the contribution to Eq. (\ref{f0})
from the mirror $D$-particles that result from the lattice 
translations.  Therefore, Eq. (\ref{f0}) changes into
\begin{eqnarray}
 f_0 & = & 1 + g_s l_s^7 \sum_{n=1}^{N} \sum_{n_{10-p} = 
-\infty}^{\infty} \cdots \sum_{n_9 = -\infty}^{\infty} \nonumber \\
 & & \times \frac{Q_n}{ \left( (x_{10-p} - x_{n(10-p)} + 
n_{10-p} R_{10-p} )^2 + \cdots (x_9 - x_{n9} + n_9 R_9 )^2 + 
| \vec{r} - \vec{r}_n |^2 \right)^{7/2} }
\label{tempf}
\end{eqnarray}
where we introduce $\vec{r} = (x_1, \ \cdots \ , x_{9-p} )$.  
In the decompactified limit where $R_i \rightarrow \infty$,
the leading contribution to Eq. (\ref{tempf}) comes from the 
case when $n_{10-p} =  \cdots  = n_9 =0$, yielding Eq. (\ref{f0}).
On the other hand, for a vanishingly small torus, i.e., in the 
$R_i \rightarrow 0$ limit, we can replace the summations in 
Eq. (\ref{tempf}) with integrations.  We can thus approximate
\begin{eqnarray}
f_0 & \simeq & 1 + g_s l_s^7 \sum_{n=1}^{N} 
\frac{Q_n}{R_{10-p} \cdots R_9}
\int_{-\infty}^{\infty} \cdots \int_{-\infty}^{\infty}   
dt_{10-p} \cdots dt_9 \nonumber \\
 & & \times \frac{1}{ \left( (x_{10-p} - x_{n(10-p)} + 
t_{10-p} )^2 + \cdots +
(x_9 - x_{n9} + t_9 )^2 + | \vec{r} - \vec{r}_n |^2 \right)^{7/2} } 
                 \nonumber \\
 & = &  1 + \frac{(5-p)!!}{15} 
    \frac{\pi^p g_s l_s^7}{R_{10-p} \cdots R_9} 
\sum_{n=1}^{N} 
\frac{Q_n}{| \vec{r} - \vec{r}_n |^{7 - p} }  ,
\label{intf}
\end{eqnarray}
where we explicitly evaluate the integrations going from the first
line to the second.   We note that, upon the integration over $t_i$s,
the translational invariance of the $t_i$ variables makes $f_0$
independent of $x_{10-p}, \ \cdots $, and $x_9$.  In other words, due to its 
small size, we can no longer resolve the position on the torus
$T^p$ at the level of the classical metric.  Eq. (\ref{intf})
is sensible for $p = 1, \ \cdots \ , 6$ and $p =8$.  However, the
double factorial function $(n)!!$ has a simple pole for $n = -2$,
and we have to replace Eq. (\ref{intf}) with a logarithmic
function for $p=7$.  

Since we take the limit $R_i \rightarrow 0$ to obtain Eq. (\ref{intf}), 
it is natural to take $T$ dual transformation
to each circle of the $T^p$.  Upon taking the $T$ dual transformations,
the metric expression becomes
\begin{equation}
ds^2_{10} =  \frac{1}{\sqrt{f_0}} ( - dt^2 + dx_{10-p}^2 + \ \cdots \ 
+ dx_9^2 ) + \sqrt{f_0} (dx_1^2 + \  \cdots \  + dx_{9-p}^2  ),
\end{equation}
while the dilaton changes to
\begin{equation}
e^{-2 \phi} = g_s^{\prime -2} f_0^{(p-3) /2} ,
\end{equation}
which naturally gives $D$ $p$-brane solutions on the dual torus with
radii $R^{\prime}_i = l_s^2 / R_i$.  
The dual string coupling $g_s^{\prime}$
is related to the original string coupling $g_s$ via 
\[ g_s = g_s^{\prime} 
   \frac{l_s^p}{R^{\prime}_{10-p} \cdots R^{\prime}_9} . \]
 The function $f_0$, in terms of dual string coupling, can be written as
\[ f_0 =  1 + \frac{(5-p)!!}{15} \pi^p g^{\prime}_s l_s^{7-p} 
\sum_{n=1}^{N} \frac{Q_n}{| \vec{r} - \vec{r}_n |^{7 - p} } , \]
which is the well-known harmonic function for extremal $D$ $p$-branes.
For a finite $g^{\prime}_s$, the function $f_0$ is regular 
under the limit $R_i \rightarrow 0$.

\subsection{Infinite Lorentz boost along the $M$ theory circle
toward DLCQ $M$ theory}

In this subsection, we lift the ten-dimensional $D$-particle solutions
compactified on the $T^p$ into the eleven dimensions by the type IIA/$M$ 
theory connection.  
Following the prescription of Seiberg \cite{seiberg}, 
we take the infinite boost along the $M$ theory circle.  
For this purpose, it is more convenient to rewrite the IIA 
parameters $g_s$ and $l_s$ in terms of the eleven dimensional
supergravity parameters $R_s$, the circumference of the $M$ 
theory circle, and $l_p$, the eleven dimensional Planck scale.  
They are related to each other by $g_s = (R_s / l_p )^{3/2}$ 
and $l_s^2 = l_p^3 / R_s$.  The function
$f_0$ of Eq. (\ref{tempf}) can be rewritten as
\begin{eqnarray}
 f_0 & = & 1 + \frac{l_p^9}{R_s^2} 
\sum_{n=1}^{N} \sum_{n_{10-p} = -\infty}^{\infty} \cdots 
\sum_{n_9 = -\infty}^{\infty} \nonumber \\
& & \times \frac{Q_n}{ \left( (x_{10-p} - x_{n(10-p)} 
+ n_{10-p} R_{10-p} )^2 + \cdots
(x_9 - x_{n9} + n_9 R_9 )^2 + | \vec{r} - \vec{r}_n |^2 \right)^{7/2} }
\label{10f}
\end{eqnarray}
The eleven-dimensional supergravity metric can be computed using
Eq. (\ref{10f}) and Eq. (\ref{rr}) to give
\begin{eqnarray}
 ds_{11}^2 & = & g_s^{2/3} e^{-2 \phi /3 } ds_{10}^2  +
g_s^{-4/3}  e^{4 \phi /3} (dx_{11} - 
A_t dt)^2  \nonumber \\
& = & -\frac{1}{f_0} dt^2 + f_0 \left(dx_{11} - ( 1- \frac{1}{f_0} ) dt 
\right)^2 + dx_1^2 + \ \cdots \ + dx_9^2 .
\label{11temp}
\end{eqnarray}
According to the dual description between the type IIA theory
and the $M$ theory on a circle, the eleventh spatial coordinate
$x_{11}$ has the period of $R_s$ and thus should be identified
via the lattice translation
\begin{equation}
x_{11} \simeq x_{11} + R_s \ , \ t \simeq t .
\end{equation}
The key observation is that in terms of light-cone coordinates
$x^{\pm} = x_{11} \pm t$, the metric Eq. (\ref{11temp}) can be
written as
\begin{equation}
 ds_{11}^2 = dx^+ dx^- + h_0 dx^- dx^- + dx_1^2 + \ \cdots \ 
dx_9^2 
\label{lc}
\end{equation}
while the lattice translation becomes
\begin{equation}
x^+ \simeq x^+ + R_s \ , \ x^- \simeq x^- + R_s  ,
\label{lci}
\end{equation}
where the function 
\begin{equation}
h_0 = f_0 -1 .
\end{equation}
We note that Eq. (\ref{lc}) and Eq. (\ref{lci}) are identical
to the equations considered in Ref. \cite{hsk}.  Under 
the infinite Lorentz boost by the boost parameter
\[ \beta = \frac{R}{\sqrt{R^2 + 4 R_s^2 } } \]
taking the limit $\beta \rightarrow 1 $ and $R_s \rightarrow 0$
while keeping $R$ fixed, the lattice translation approaches
\[ x^- \simeq x^- + R \ , \ x^+ \simeq x^+  . \]
In other words, $M$ theory on the original spatial circle with 
the size $R_s$ turns into the $M$ theory on a light-cone
circle with the size.  
In this framework, $x^+$ plays the role of a 
time coordinate, and $x^-$ plays the role of the eleventh
coordinate.  We note that the effect of the boost for the metric 
expression results the substitution
\begin{eqnarray}
 h_0 & \rightarrow & h = \frac{R_s^2}{R^2} h_0 
= \frac{l_p^9}{R^2} 
\sum_{n=1}^{N} \sum_{n_{10-p} = -\infty}^{\infty} \cdots 
\sum_{n_9 = -\infty}^{\infty} \nonumber \\
& & \hspace{2cm} \times \frac{Q_n}{ \left( 
(x_{10-p} - x_{n(10-p)} + n_{10-p} R_{10-p} )^2 + \cdots
(x_9 - x_{n9} + n_9 R_9 )^2 + 
| \vec{r} - \vec{r}_n |^2 \right)^{7/2} }
\label{hafterb}
\end{eqnarray}
and the eleven-dimensional metric after the boost is
\begin{equation}
 ds_{11}^2 = dx^+ dx^- + h dx^- dx^- + dx_1^2 + \ \cdots \ 
dx_9^2 .
\label{afterb}
\end{equation}
The non-vanishing $h$, which is proportional to the $D$-particle 
numbers, provides a natural regulator that makes $x^+$ a true time 
coordinate and $x^-$ a true spatial coordinate at the level of the 
classical metric. Strictly speaking, the metric Eq. (\ref{afterb}) is 
valid for small value of $R_s$ (and $R$), since it is the 
dimensionally lifted version from the ten-dimensional solutions.  
During the dimensional 
reduction, the massive Kaluza-Klein modes disappear, and the 
contribution from these modes do not show up in Eq. (\ref{afterb}).
In subsections that follow, we will find the generalization of
Eq. (\ref{afterb}) that is valid for a finite $R$.  In fact,
since we will ultimately be interested in the case when 
$R \rightarrow \infty$, it is essential to include the massive
Kaluza-Klein mode contribution, for they become massless in that
limit.

Upon the compactification on the $x^-$ circle, Eq. (\ref{afterb})
reduces to non-asymptotically flat solutions of the type IIA 
supergravity \cite{hsk}.  An interesting point is that the
same non-asymptotically flat solutions are obtained by starting
from the usual $D$-particle (or $D$ $p$-brane in the $T$-dualized
version) solutions and concentrating on {\em near-horizon}
geometry \cite{malda}.  We will elaborate on this point in the later
subsections.
   
\subsection{$T^0$ : BFSS matrix theory}

We start from the case originally considered by Banks, Fischler,
Shenker and Susskind in their formulation of the matrix
theory \cite{bfss}.  The eleven dimensional metric in this case is 
Eq. (\ref{afterb}) with $p=0$
\begin{equation}
 ds_{11}^2 = dx^+ dx^- + \frac{l_p^9}{R^2}
\sum_{n=1}^N \frac{Q_n}{|\vec{r} - \vec{r}_n |^7 }  dx^- dx^- 
+ dx_1^2 + \ \cdots \ + dx_9^2 .
\label{ini0}
\end{equation}
This metric describes the space-time geometry for small $R$,
since it only contains the contribution
from the Kaluza-Klein zero mode.  To motivate the metric expression 
that is valid for a finite $R$, we note that Eq. (\ref{ini0})
(with $N=1$) is the Kaluza-Klein zero mode part ($1/R$ part of
the delta function $\delta(x^- )$) of the eleven 
dimensional Aichelberg-Sexl metric \cite{aich}
\begin{equation}
ds_{11}^2 = dx^+ dx^- + l_p^9 \frac{ p_-}{r^7} \delta(x^- ) 
dx^- dx^- + dx_1^2 + \ \cdots \ + dx_9^2 
\label{ase}
\end{equation}
representing the graviton moving along the asymptotic light-cone 
direction, where we identify the light-cone momentum 
$p_- = Q_1 / R $ \cite{becker}. To include the contribution to the metric 
from all the massive Kaluza-Klein modes, we generalize 
Eq. (\ref{ini0}) as follows.
\begin{equation}
 ds_{11}^2 = dx^+ dx^- + l_p^9 \sum_{n=1}^N 
\sum_{n_0 = -\infty}^{\infty} 
\frac{p_{-n}  \delta ( x^- - x_n^- + n_0 R ) }
{|\vec{r} - \vec{r}_n |^7 }    
dx^- dx^-  + dx_1^2 + \ \cdots \ + dx_9^2 ,
\label{fini0}
\end{equation}
where we introduce two sets of parameters that are used to
denote the light-cone momentum $p_{-n} = Q_n / R$ and the
light-cone position $x_n^-$.  Eq. (\ref{fini0}) is manifestly
invariant under the lattice translation $x^- \rightarrow x^- 
+ R$, which is reminiscent of the requirement of the modular 
invariance for the string amplitudes.  Furthermore, since the 
light-cone circle size is finite, we should be able to resolve the 
position of the graviton $x_n^-$ at the level of the classical
metric.  

As we take the limit $R \rightarrow 0$,
we can replace the infinite summation in Eq. (\ref{fini0}) into an 
integral to recover Eq. (\ref{ini0}).  To this effect, we note that
\[ \sum_{n_0 = - \infty}^{\infty}  \delta ( x^- - x_n^- + n_0 R   )
\ \rightarrow \ \frac{1}{R} \int_{-\infty}^{\infty} dt_0 
\delta (  x^- - x_n^- + t_0  ) = \frac{1}{R} . \]
As in the subsection A, the information of the position of the
graviton encoded in the metric disappears in this limit due to
the appearance of the continuous translational symmetry.

To understand the background geometry of the DLCQ $M$ theory,
we are interested in the limit where $R \rightarrow \infty$.
Following \cite{susskind}, we let the largest value among 
$Q_n$'s grow proportional to $R$ to give a fixed light-cone 
momentum $p_-$\footnote{Alternatively, we can set 
$\vec{r}_1 = \cdots = \vec{r}_N = \vec{r}_{n_{max}} $ and
$x_1^- = \cdots = x_N^- = \vec{x}_{n_{max}}^-$ by concentrating
on single-center solutions.  In this case, we define
$Q_{n_{max}} = Q_1 + \cdots + Q_N$ and the argument
proceeds with no further corrections.  The same remark also
appliles to the following subsections.}.  
In this limit, the leading contribution
to Eq. (\ref{fini0}) comes from $n_0 = 0$ and $n = n_{max}$
such that $Q_{n_{max}}$ is the largest value among $Q_n$'s.
By setting $p_- = Q_{n_{max}} / R $, the metric expression
becomes
\begin{equation}
ds_{11}^2 = dx^+ dx^- + l_p^9  
\frac{p_-  \delta ( x^- - x_0^-  ) }
{r^7 }    
dx^- dx^-  + dx_1^2 + \ \cdots \ + dx_9^2 ,
\label{0back}
\end{equation}
where $x_0^- = x_{n_{max}}^-$ and we introduce
$r^2 = | \vec{r} - \vec{r}_{n_{max}} |^2$.  
We see from Eq. (\ref{0back})
that $M$ momentum effect is localized at $x^- = x_0^-$,
and we call this behavior `light-cone momentum localization'.
For $x^- \ne x_0^-$, the metric Eq. (\ref{0back}) becomes
the eleven dimensional Minkowski space-time $R^{1,10}$.
Precisely at the support of the delta function ($x^- = x_0^-$),
the metric becomes
\begin{equation}
ds_{x_0^-}^2 = dx_1^2 + \ \cdots \ + dx_9^2 
\end{equation}
which is $R^9$.  Since $x^{\pm}$ are light-cone variables,
the first two terms in Eq. (\ref{0back}) vanish at a fixed $x^-$.
To summarize, the background geometry of $M$ theory is
$R^{1,10}$, and the $D$-particles, or $M$ momentum, that play the 
role of partons in BFSS theory, are localized in 
the $R^9$ submanifold located at $x^- = x_0^-$.  This is
consistent with BFSS theory where we write down the 
D-particle hamiltonian, and $D$-particles move on the
target space of the transversal $R^9$, thereby giving the
Hilbert space of $N$-symmetric product of $R^9$, i.e., $R^{9N} / S_N$.  
The microscopic description is holographically localized into the 
transversal $R^9$.  We note that $R^9$ at $x^- = x_0^-$ plays the 
role of $R^{1, d-1}$ at the spatial infinity of AdS$_{d+1}$ in the
context of the AdS/CFT correspondence. 

It is instructive to notice the parallel between the holographic
description in the above and the Seiberg's description of the
microscopic theory in terms of $\tilde{M}$ theory on a spatial circle 
with the vanishingly small radius $\tilde{R}$.  For a typical
characteristic transversal scale $\tilde{R}_i$ of $\tilde{M}$ 
theory we let
\[ \frac{R_i}{l_p} = \frac{\tilde{R}_i}{\tilde{l}_p } , \]
and the rescaled eleven-dimensional Planck length is required
to satisfy
\[ \frac{R}{l_p^2} = \frac{\tilde{R}}{\tilde{l}_p^2 } \]
to give physically the same description as the DLCQ $M$ theory.
In $\tilde{M}$ theory, the string coupling, the string length,
the size of the eleventh spatial circle, and the characteristic 
transversal length scale is given by 
$\tilde{g}_s = (\tilde{R} / R )^{3/4} g_s $,
$\tilde{l}_s = (\tilde{R} / R )^{1/4} l_s $,
$\tilde{g}_s \tilde{l}_s = \tilde{R}$, and
$\tilde{R}_i = (\tilde{R} / R )^{1/2} R_i $.  Due to the 
vanishing $\tilde{M}$ theory circle size in the limit 
$\tilde{R} \rightarrow 0$ while keeping $M$ theory parameters
fixed, the $M$ momentum of $M$ theory becomes $D$-particle
number of $\tilde{M}$ theory.  Furthermore, 
$\tilde{g}_s , \tilde{l}_s \rightarrow 0$ in this limit, freezing
the bulk dynamics and the massive string excitations.  We also
note that the size of the eleventh circle vanishes infinitely 
faster than the transversal length scale to yield $R^9$, which
corresponds to the spatial parts of the ten-dimensional
IIA theory.  These are the behaviors that we also observed in
terms of the background geometry of the DLCQ $M$ theory.

\subsection{$T^1$ : DVV matrix string theory}

$M$ theory compactified on a circle, ninth spatial direction with
a vanishingly small circumference $R_9$, was given a successful
matrix string theory formulation by Dijkgraaf, Verlinde and
Verlinde \cite{dvv}.  To discuss the background geometry of
this formulation, we start from the small $R$ and $R_9$ expressions
Eqs. (\ref{hafterb}) and (\ref{afterb}).  In the end, we will
be interested in the limit $R \rightarrow \infty$, as in the
previous subsection.  After replacing the summation over $n_9$ with an 
integration along the fashion of Eq. (\ref{intf}) and 
dimensionally-reducing it to ten dimensions along the light-cone eleventh 
circle $x^-$, we get the following IIA supergravity metric, the dilaton 
field, and RR one-form gauge field.
\begin{equation}
ds^2_{10} = - \frac{1}{\sqrt{h}} d \tau^2 + \sqrt{h} 
    (dx_1^2 + \  \cdots \  + dx_9^2  ),
\label{t1metric}
\end{equation}
\[ e^{- 2 \phi} = \left( \frac{R^{3/2}}{l_p^{3/2}} 
      \right)^{-2} h^{-3/2} \ \ , \ \
   A_{\tau} = - \frac{1}{h}  \]
Here $h$ is a eight-dimensional harmonic function
\begin{equation}
 h = \frac{8 \pi}{15} \frac{l_p^9}{R^2 R_9}
\sum_{n=1}^{N} \frac{Q_n}{| \vec{r} - \vec{r}_n |^6}  , 
\label{t1h}
\end{equation}
where $\vec{r}$ is an eight-dimensional vector.  In addition,
we introduced a time coordinate $ \tau  =  x^+ /2 $.  We note that
$x_1, \cdots , x_8$ are non-compact coordinates, and $x_9$ denotes
the circumference position of the ninth spatial circle with the
circumference $R_9$.  The metric Eq. (\ref{t1metric}), that is
non-asymptotically flat in the ten-dimensional sense, is precisely
that of the large $Q_n$ and/or near-horizon metric of the usual
$D$-particle solutions compactified on a spatial circle.   

The key insight from Ref. \cite{dvv} is that we can achieve
nine-eleven flip via the chain of $TST$ dualities where
the $T$-dual transformation acts on the ninth circle.  
Under this chain of dual transformations, the original $D$-particle
number changes to $D$-string winding number, then to
fundamental string winding number, and finally to the 
fundamental string momentum.  In addition, we will see
from Eqs. (\ref{flipmetric}) and (\ref{11flip}) below
that $TST$ transformation renders the space-time geometry
asymptotically-flat while making the dilaton field a constant.
In fact, after directly applying this chain of dualities to 
Eq. (\ref{t1metric}), we find 
that the non-vanishing IIA supergravity fields are the metric
\begin{equation}
ds_{10}^2 =  d x^+ dx_9 + h dx_9^2 + dx_1^2 + \cdots 
 + dx_8^2 
\label{flipmetric}
\end{equation}
and the dilaton
\begin{equation}
e^{-2 \phi}  = \left( \frac{R_9^{3/2} }{ l_p^{3/2} } \right)^{-2} ,
\label{flipdilaton}
\end{equation}
where we use the same $h$ as in Eq. (\ref{t1h}) and we use
$x^+$ in favor of $\tau =  x^+ / 2 $.  The chain of dualities 
change the circumference of the ninth circle into $R$, the original
$M$ theory circle circumference (see the table 2 below).  Furthermore, 
as Eq. (\ref{flipmetric}) shows, the ninth circle, which was originally
a spatial circle when $h=0$, is now a light-like circle when $h=0$.
By lifting Eqs. (\ref{flipmetric}) and (\ref{flipdilaton}) into
the eleven dimensions, we find that the eleven dimensional
metric now reads
\begin{equation}
ds_{11}^2 = dx^+ dx_9 + h dx_9^2 + dx_1^2 + 
\cdots + dx_8^2 + dx^- dx^-  .
\label{11flip}
\end{equation}
This equation shows that the original light-like circle (when $h=0$)
$x^-$ turns into a spatial circle with the circumference
$R_9$ (see the table 2 below) under $TST$ dualities.  
Compared to Eq. (\ref{ini0}), we directly observe that
the eleven dimensional metric Eq. (\ref{11flip}) is precisely
the nine-eleven flipped version of Eq. (\ref{ini0}).  The only
difference is that while the coefficient of $dx^- dx^-$ term in
Eq. (\ref{ini0}) is a nine-dimensional harmonic function 
(produced by a point-like $D$-particle source in the language of 
IIA string theory), 
the coefficient $h$ of $dx_9^2$ in Eq. (\ref{11flip}) is
a eight-dimensional ($x_1 , \cdots , x_8 $) harmonic function.  
The source of $h$ is of course fundamental strings (one-dimensional source) 
with zero winding number along the $x^9$ light-cone circle in IIA string 
theory.  In the language of eleven-dimensional
supergravity, the metric Eq. (\ref{11flip})
describes the space-time geometry produced by the momentum
of the longitudinal membranes (extended over $(x_9 , x^- )$) wrapped
along the $M$ theory spatial circle (but not wrapped along the $x_9$
circle).  Since these membranes do not wind around the full $T^2$,
they do not produce electric charges for the three-form gauge field.

To better understand the $TST$ transformations that lead us to
Eq. (\ref{11flip}), we observe that the string coupling, string length
scale, the ninth circle circumference, and the eleventh circle 
circumference change as follows under each duality transformation.
\\
\begin{center}
\begin{tabular}{|c|c|c|c|c|} \hline
 Type  & string coupling  & string length scale & ninth circle  & 
         eleventh circle \\ 
       &   &   & circumference & circumference   \\ \hline 
  IIA  & $g_s \equiv (R/ l_p )^{3/2}$ & $l_s \equiv
         (l_p^3 / R )^{1/2}$ & $R_9$  & $g_s l_s = R$  \\ \hline
  IIB  & $g_s l_s / R_9 $ & $l_s$ & $l_s^2 / R_9 $ &   \\ \hline
  IIB  & $R_9 / (g_s l_s )$ & $g_s^{1/2} l_s^{3/2} / R_9^{1/2} $ 
       & $ l_s^2 / R_9 $  & \\ \hline
  IIA  & $R_9^{3/2}/(g_s^{1/2} l_s^{3/2} ) = ( R_9 / l_p )^{3/2}$ & 
         $g_s^{1/2} l_s^{3/2} / R_9^{1/2} = ( l_p^3 / R_9 )^{1/2} $ 
       & $g_s l_s = R $  & $R_9$ \\ \hline
\end{tabular}
\\
[.3cm]
 Table 2. We take $T$, $S$ and $T$ dualities, successively.  
\end{center}
Under $T$ duality along the ninth direction, the parameters
transform like $R_9 \rightarrow l_s^2 / R_9 $, 
$g_s \rightarrow g_s l_s / R_9$, and $l_s \rightarrow l_s$.
On the other hand, under $S$ duality, the parameters transform like
 $R_9 \rightarrow R_9 $, $g_s \rightarrow g_s^{-1} $, and 
$l_s \rightarrow \sqrt{g_s} l_s$.  The eleven-dimensional Planck
length $l_p$ remains invariant under the $TST$ transformations. 

To investigate the behavior of $M$ theory on $T^1$, we have to take
the $R \rightarrow \infty$ limit while keeping $R_9$
vanishingly small.  Just like the analysis of the $T^0$ case, 
the massive Kaluza-Klein momentum modes along the $x_9$ circle give
additional contributions to the metric expression.  We thus write
the metric expression as
\begin{equation}
ds_{10}^2 = dx^+ dx_9 +  \frac{8 \pi}{15} \frac{l_p^9}{R_9}
\frac{p_9 \delta ( x_9 - x_{90} ) }{r^6} dx_9^2 
+ dx_1^2 + \cdots + dx_8^2 .
\label{t1back}
\end{equation}
We drop the eleventh circle since it is vanishingly small,
and we set $p_9 = Q_{n_{max}} / R$ and $r^2 =
| \vec{r} - \vec{r}_n |^2$.  
We note that, in terms of the $TST$-transformed string coupling
$g_s^{\prime}$ and the string length scale $l_s^{\prime}$, we have
$l_p^9 / R_9 = g_s^{\prime 2} l_s^{\prime 8}$, which is the same
as the ten-dimensional Newton's constant up to a numerical factor.
The parameter $x_{90}$ denotes
the longitudinal center of mass position of the string.  The background
geometry is the ten-dimensional Minkowski space-time $R^{1,9}$
as far as $x_9 \ne x_{90}$.  The effect of the string momentum
is holographically localized at $x_9 = x_{90}$; precisely at the 
support of the delta function $x_0 = x_{90}$, the first two terms of 
Eq. (\ref{t1back}) vanish to yield the metric localized
at the support
\begin{equation}
ds_{x_{90}}^2 = dx_1^2 + \cdots + dx_8^2 ,
\end{equation}
which is $R^8$, the transversal space of the string.  This is 
consistent with DVV theory, where strings move on the target
space of the transversal $R^8$, thereby giving the Hilbert
space of $R^{8N} / S_N$.  

Via $\tilde{M}$ theory of Seiberg, we can understand the microscopic
theory better.  We note that under the $TST$ transformations
the parameters of the $\tilde{M}$ theory transforms as follows.    
\\
\begin{center}
\begin{tabular}{|c|c|c|c|c|} \hline
 Type  & string coupling  & string length scale & ninth circle  & 
         eleventh circle  \\ 
       &   &   & circumference & circumference  \\ \hline 
  IIA  & $ \left( \frac{\tilde{R}}{R} \right)^{3/4} g_s $ 
       & $ \left( \frac{\tilde{R}}{R} \right)^{1/4} l_s $
       & $ \left( \frac{\tilde{R}}{R} \right)^{1/2} R_9 $  
       & $ \tilde{R} $  \\ \hline
  IIB  & $ \left( \frac{\tilde{R}}{R} \right)^{1/2} g_s l_s / R_9 $ 
       & $ \left( \frac{\tilde{R}}{R} \right)^{1/4} l_s$ 
       & $l_s^2 / R_9 $ 
       &        \\ \hline
  IIB  & $ \left( \frac{R}{\tilde{R}} \right)^{1/2} R_9 / (g_s l_s )$ 
       & $ \left( \frac{\tilde{R}}{R} \right)^{1/2} 
           g_s^{1/2} l_s^{3/2} / R_9^{1/2} $ 
       & $ l_s^2 / R_9 $  
       &        \\ \hline
  IIA  & $( R_9 / l_p )^{3/2}$ 
       & $ \left( \frac{\tilde{R}}{R} \right)^{1/2} ( l_p^3 / R_9 )^{1/2} $ 
       & $\tilde{R} $  
       & $ \left( \frac{\tilde{R}}{R} \right)^{1/2} R_9$ \\ \hline
\end{tabular}
\\
[.3cm]
 Table 3. We take $T$, $S$ and $T$ dualities in $\tilde{M}$ theory, 
         successively.  
\end{center}
The characteristic length scale of the non-compact direction satisfies
$\tilde{R}_i = R_i (\tilde{R}/R )^{1/2}$ where $i=1, \cdots , 8$, 
and $\tilde{l}_p = l_p (\tilde{R}/R )^{1/2}$.
We mention that both the ninth circle and the eleventh circle
are the spatial circles in $\tilde{M}$ theory. 
The second row was the one analyzed by Seiberg \cite{seiberg}.  
The fourth row, the DVV theory case, is also interesting, for
the string mass scale that is the inverse of the string length
scale diverges as we take the limit $\tilde{R} \rightarrow 0$,
just like the second row.  This implies that the massive string
excitations freeze.  Likewise, since the string coupling
remains finite under the $\tilde{R} \rightarrow 0$ limit
(showing that the basic objects of this theory are strings)
and becomes very small for small $R_9$,
the bulk dynamics and the string loop effects can also be
neglected, just like the case of the second row.  On the contrary,
in case of the third row, the string coupling diverges in the
limit $\tilde{R} \rightarrow 0$, making it virtually impossible
to write down the fundamental hamiltonian.  The main difference between
the second row and the fourth row is the size of the 
ninth circle.  In the case of the second row, the characteristic
transversal length scale vanishes infinitely faster than the ninth 
circle circumference (that becomes finite in the limit 
$\tilde{R} \rightarrow \infty$ and becomes large for a small
value of $R_9$).  Thus, the microscopic description in this 
case is (1+1)-dimensional supersymmetric Yang-Mills theory 
where the $D$-string numbers play the role of parton numbers.  This is
inherently the IIB type matrix theory.  In the case of the fourth row,
however, which gives inherently the IIA type matrix string theory, 
the ninth circle circumference vanishes infinitely faster
than the non-compact characteristic length scale and the eleventh 
circle circumference, as we take the limit 
$\tilde{R} \rightarrow 0$.  This is consistent with our background
metric Eq. (\ref{t1back}) where localizing $x_9 = x_{90}$ 
decouples the ninth (light-cone) circle.  Furthermore, similar 
to what happens
to the background metric, the circumference of the eleventh
circle, that scales as $\tilde{R}^{1/2}$ like the transversal
characteristic length scale, becomes very small for a small
value of $R_9$ relative to the transversal length scale.  Thus, it is 
natural to apply the prescription
of Tayler \cite{wat} to fourier-decompose ($T$-dualize) the small 
eleventh circle getting the DVV hamiltonian from the BFSS hamiltonian.
In both $T^0$ and $T^1$ cases, the background geometry
is flat Minkowskian, the geometry localized at the light-cone 
momentum is flat Euclidean ($R^8$ or $R^9$), and the parton-like
object is the light-cone momentum.      
 
\subsection{$T^2$ : AdS$_4$/CFT$_3$ Correspondence}

The compactification of DLCQ $M$ theory on $T^2$ provides
us with the holographic description
via AdS/CFT correspondence.  We start from the case when
$R$, $R_8$ and $R_9$ are small.  The dimensional
reduction along the light-cone $M$ theory circle 
of Eqs. (\ref{hafterb}) and (\ref{afterb}) yields the following 
ten-dimensional IIA supergravity fields.
\begin{equation}
ds^2_{10} = - \frac{1}{\sqrt{h}} d \tau^2 + \sqrt{h} 
    (dx_1^2 + \  \cdots \  + dx_9^2  ),
\label{t2metric}
\end{equation}
\[ e^{- 2 \phi} = \left( \frac{R^{3/2}}{l_p^{3/2}} 
      \right)^{-2} h^{-3/2} \ \ , \ \
   A_{\tau} = - \frac{1}{h}  \]
Here $h$ is a seven-dimensional harmonic function
\begin{equation}
 h = \frac{\pi^2}{5} \frac{l_p^9}{R^2 R_8 R_9}
\sum_{n=1}^{N} \frac{Q_n}{| \vec{r} - \vec{r}_n |^5}  , 
\label{t2h}
\end{equation}
where $\vec{r}$ is a seven-dimensional vector 
representing $(x_1 , \cdots , x_7 )$, the non-compact
part of the space-time.  Since $R_8$ and $R_9$ are
small, it is natural to take $T_8 T_9$ duality transformation.
This transformation renders the metric and the dilaton
field into
\begin{equation}
ds^2_{10} =  \frac{1}{\sqrt{h}} ( - d \tau^2 +  dx_8^2 + dx_9^2
) + \sqrt{h} (dx_1^2 + \  \cdots \  + dx_7^2  ),
\label{d2metric}
\end{equation}
\[ e^{- 2 \phi} = \left( \frac{R^{1/2} l_p^{3/2}}{R_8 R_9} 
      \right)^{-2} h^{-1/2} .  \] 
The only non-vanishing RR gauge field is the RR three-form 
electric gauge field.  The metric Eq. (\ref{d2metric}) is
actually identical to the {\em near-horizon} metric of the
$D$-membrane solutions and, as such, it is non-asymptotically
flat in the ten-dimensional sense.  

The $T_8 T_9$ transformation keeps the ten-dimensional
string length scale fixed.  However, the string coupling constant
changes into $g_s^{\prime} = g_s l_s^2 / (R_8 R_9 )$, and the
radii of the eighth and the ninth circle become
$R_8^{\prime} = l_s^2 / R_8 $ and $R_9^{\prime} = l_s^2 / R_9$,
respectively.  These changes in the parameters translate
to the transformed eleven-dimensional Planck length 
$l_p^{\prime} = l_p^2 / (R R_8 R_9 )^{1/3}$ and the
eleventh circumference $R_{11}^{\prime} = l_p^3 / (R_8 R_9 )$.
We note that
\begin{equation}
   \frac{R_{11}^{\prime}}{l_p^{\prime}}
   = \left( \frac{l_p^{3/2} R^{1/2}}{R_8 R_9} \right)^{2/3}  . 
\label{crite}
\end{equation}
We started from a vanishingly small values of $R_8$
($R_9$) and, for this purpose, we require that the original 
$R_8$ ($R_9$) should be kept smaller than the string length 
scale
\begin{equation}
 R_8 \ll l_s = \left( \frac{l_p}{R} \right)^{1/2} l_p  \ , \
 R_9 \ll l_s = \left( \frac{l_p}{R} \right)^{1/2} l_p
\label{cond1}
\end{equation}
regardless of the $R/l_p$ ratio.  Thus, the dual sizes
$R_8^{\prime}$ and $R_9^{\prime}$ are infinitely large
compared to the string length scale.  Furthermore, the 
conditions Eq. (\ref{cond1}) imply that
\begin{equation}
   \frac{R_{11}^{\prime}}{l_p^{\prime}}
   = \left( \frac{l_p^{3/2} R^{1/2}}{R_8 R_9} \right)^{2/3} 
  \gg \frac{R}{l_p} . 
\label{cond2}
\end{equation}
This equation shows that the eleventh direction naturally
decompactifies in the limit where the original light-cone circle 
decompactifies ($R \gg l_p$); the background geometry
in this limit is eleven-dimensional.  

Since the RR one-form gauge field identically vanishes, the 
dimensionally lifted version of Eq. (\ref{d2metric}) into
eleven dimensions via Eq. (\ref{11temp}) determines the
background metric to be
\begin{equation}
ds_{11}^2 = h^{-2/3} (- d\tau^2 + dx_8^2 + dx_9^2 )
  + h^{1/3} ( dx_1^2 + \cdots + dx_7^2 + dx_{11}^2 ) ,
\label{t2temp}
\end{equation}
where the function $h$ is rewritten as
\begin{equation}
  h = \frac{8 \pi}{15} l_p^{\prime 6}
\left( \frac{3 \pi}{8} \frac{1}{R_{11}^{\prime}} 
 \sum_{n=1}^{N} \frac{Q_n}{|\vec{r} - \vec{r}_n |^5} \right)
\label{2h}
\end{equation}
in terms of the transformed eleven dimensional parameters.
Eq. ({\ref{t2temp}) is valid as long as $R^{\prime}_{11}$ is
vanishingly small, and it is the same as the near horizon metric 
of $M$-membranes where the $x_{11}$ direction transversal to
the membrane is compactified to a small radius (by setting $R$ 
infinitesimally small).  When the eleventh direction is 
compactified with a finite radius, we have to add the
contributions from the mirror membranes that result from
the lattice translation \cite{imsy}.  Thus, Eq. (\ref{2h}) can 
be generalized to
\begin{equation}
 h = \frac{8 \pi}{15} l_p^{\prime 6}
\sum_{n=1}^{N} \sum_{n_0 = -\infty}^{\infty} \frac{Q_n}
{\left( |\vec{r} - \vec{r}_n |^2 + 
   (x_{11} - x_{n11} + n_0 R_{11}^{\prime} )^2 \right)^3} , 
\end{equation}
which, in the limit $R_{11}^{\prime} \rightarrow 0$ where
we can place the $n_0$ summation with an integration,
reduces to Eq. (\ref{2h}) as follows.
\begin{eqnarray}
 & &  \sum_{n_0 = -\infty}^{\infty} \frac{Q_n}
{\left( |\vec{r} - \vec{r}_n |^2 +  
   (x_{11} - x_{n11} + n_0 R_{11}^{\prime} )^2 \right)^3} 
      \nonumber \\
 &  \rightarrow  & \frac{1}{R_{11}^{\prime}} 
    \int_{-\infty}^{\infty}  
 \frac{Q_n dt } {\left( |\vec{r} - \vec{r}_n |^2 + 
   (x_{11} - x_{n11} + t )^2 \right)^3}
 = \frac{3\pi}{8} \frac{1}{R_{11}^{\prime}}
   \frac{Q_n}{|\vec{r} - \vec{r}_n |^5}  \nonumber
\end{eqnarray}
Our primary interest is the decompactified limit
where we require $R \gg l_p$; in this case $R_{11}^{\prime}$
goes to infinity according to Eq. (\ref{cond2}).  We are 
then naturally led to concentrate on the largest value among
$Q_n$ and set $Q = Q_{n_{max}}$.  At the same time, we 
introduce $r^2 = |\vec{r} - \vec{r}_{n_{max}} |^2 + 
 (x_{11} - x_{n_{max} 11} )^2 $, the eight-dimensional
radial distance.  In this case, only $n_0 = 0$ contributes
to the $n_0$ summation.  Therefore, we have
\begin{equation}
h =  \frac{8 \pi}{15} l_p^{\prime 6} \frac{Q}{r^6} , 
\label{whew}
\end{equation}
which is an eight-dimensional harmonic function, since
the eleventh direction is fully decompactified and
can not be distinguished from $(x_1 , \cdots x_7 )$.
We note that
\[ l_p^{\prime 6} Q = \left( \frac{l_p^{\prime 9 }}
               {R_8^{\prime} R_9^{\prime} } \right)
\left( \frac{ Q R_8^{\prime} R_9^{\prime} }{l_p^{\prime 3}} 
\right) , \]
where the first factor in the curly bracket is proportional
to the nine dimensional gravitational constant and
the second factor is $Q / R$ in terms of the original
DLCQ $M$ theory parameters.  The second factor corresponds
to the energy of $Q$-times wrapped $M$ membranes when 
the usual factor one is added to the function $h$. 
In terms of a new radial coordinate $U = \sqrt{15/(8 \pi)}
 r^2 / l_p^{\prime^3} $ and using the function $h$ in
Eq. (\ref{whew}), we can rewrite the metric expression as
\begin{equation}
ds_{11}^2 = \left( \frac{8 \pi}{15} \right)^{1/3} 
 l_p^{\prime 2} \left( \frac{U^2}{Q^{2/3}} ( -d\tau^2 + dx_8^2
+ dx_9^2 ) + Q^{1/3} ( \frac{dU^2}{4 U^2} + d\Omega_{(7)}^2 )
\right)
\label{ads4}
\end{equation}
where $d\Omega_{(7)}^2$ is the metric on the unit seven-sphere.
The background geometry Eq. (\ref{ads4}) is 
$AdS_4 \times S^7$ as was first shown in Ref. \cite{hyun}.

The microscopic degrees of freedom in this case can be
understood in terms of $\tilde{M}$ theory compactified
on a spatial eleventh circle.  After taking $T_8 T_9$ dual
transformations in the context of $\tilde{M}$ theory,
the string coupling and the string length scale satisfy
\[ \tilde{g}_s^{\prime} = \left( \frac{\tilde{R}}{R} \right)^{1/4}
  \frac{R^{1/2} l_p^{3/2}}{R_8 R_9 } \ , \ 
    \tilde{l}_s^{\prime} = \left( \frac{\tilde{R}}{R} \right)^{1/4}
  \frac{l_p^{3/2}}{R^{1/2} } , \]
which vanishes as we take the limit $\tilde{R} \rightarrow 0$.
Both of the bulk dynamics and the massive excitation freeze.  
At the same time, the $T_8 T_9$ transformation turns the
original dynamics of $D$-particles into that of $D$-membranes.
The eighth and the ninth directional sizes of the torus
are
\begin{equation}
  \tilde{R}_8^{\prime} = \frac{l_p^3}{R_8 R} \ , \
   \tilde{R}_9^{\prime} = \frac{l_p^3}{R_9 R} 
\label{whew2}
\end{equation}
that are identical to those of $M$ theory and remain finite
under the limit $\tilde{R} \rightarrow 0$.  On the other hand,
the eleventh circle size, the eleven dimensional
Planck length, and the typical transversal length
scale ($i = 1, \cdots , 7$) are given by
\begin{equation}
   \tilde{R}_{11}^{\prime} = 
      \left( \frac{\tilde{R}}{R} \right)^{1/2}
      \frac{l_p^3}{R_8 R_9} \ \ , \ \
   \tilde{l}_p = \left( \frac{\tilde{R}}{R} \right)^{1/3}
     \frac{l_p^2}{(R R_8 R_9 )^{1/3}} \ \ , \ \
   \tilde{R}_i^{\prime} =     
      \left( \frac{\tilde{R}}{R} \right)^{1/2} R_i .  
\label{whew3}
\end{equation}
Compared to Eq. (\ref{whew2}), the length scales in Eq. 
(\ref{whew3}) vanish to zero as we take the limit $\tilde{R} 
\rightarrow 0$.  The relevant microscopic dynamics of
this theory is thus provided by the (2+1)-dimensional super
Yang-Mills theory of $D$-membranes.  We note that the
behavior of the background geometry reflects the 
scaling in Eq. (\ref{whew3}).  The eleventh size
and the transversal length scale scales the same way
(as proportional to $\tilde{R}^{1/2}$).  In the case
of the background geometry, we observed that the apparent
$SO(7)$ rotational symmetry of the background metric 
was enhanced to $SO(8)$ symmetry, since the eleventh
direction is no longer distinguishable from other
transversal directions.  

The aforementioned determination of the microscopic
theory due to Seiberg is in fact precisely that of
the CFT$_3$ side description in the context of the
holographic AdS$_4$/CFT$_3$ correspondence.  Noting that
our background geometry is $AdS_4 \times S^7$,
this consideration supplies the heuristic derivation
of the correspondence.    

\subsection{$T^3$ : AdS$_5$/CFT$_4$ Correspondence}

For the compactification on a small three-torus, it is natural to 
take $T_7 T_8 T_9$ dual transformations.  This turns the
original type IIA/$M$ theory framework into type IIB theory,
and $D$-particles are mapped into $D$ three-branes.  Unlike
the case of $T^1$ and $T^5$, the $T^p S T^p$ dual transformations
that correspond to the nine-eleven flip and supply IIA/$M$ side
formulation can not be applied in this context, for $D$ three-branes 
are self-dual under $S$ duality.  Thus, we will consider the 
IIB ten-dimensional background geometry.   

When $R$, $R_7$, $R_8$ and $R_9$ are small, the dimensional
reduction along the light-cone $M$ theory circle 
of Eqs. (\ref{hafterb}) and (\ref{afterb}) gives the following 
ten-dimensional IIA supergravity fields.
\begin{equation}
ds^2_{10} = - \frac{1}{\sqrt{h}} d \tau^2 + \sqrt{h} 
    (dx_1^2 + \  \cdots \  + dx_9^2  ),
\label{t3metric}
\end{equation}
\[ e^{- 2 \phi} = \left( \frac{R^{3/2}}{l_p^{3/2}} 
      \right)^{-2} h^{-3/2} \ \ , \ \
   A_{\tau} = - \frac{1}{h}  \]
Here $h$ is a six-dimensional harmonic function
\begin{equation}
 h = \frac{2 \pi^3}{15} \frac{l_p^9}{R^2 R_7 R_8 R_9}
\sum_{n=1}^{N} \frac{Q_n}{| \vec{r} - \vec{r}_n |^4}  , 
\label{t3h}
\end{equation}
where $\vec{r}$ is a six non-compact dimensional vector 
representing $(x_1 , \cdots , x_6 )$. Under $T_7 T_8 T_9$ duality 
transformation, the metric and the dilaton field transform to
\begin{equation}
ds^2_{10} =  \frac{1}{\sqrt{h}} ( - d \tau^2 +  dx_7^2 + dx_8^2 
+ dx_9^2 ) + \sqrt{h} (dx_1^2 + \  \cdots \  + dx_6^2  ),
\label{d3metric}
\end{equation}
\[ e^{- 2 \phi} = \left( \frac{l_p^3 }{R_7 R_8 R_9} 
      \right)^{-2}  .  \] 
The only non-vanishing RR gauge field is the RR self-dual 
four-form gauge field.  As before, the metric Eq. (\ref{d3metric}) 
is the {\em near-horizon} metric of the $D$ three-brane solutions
and non-asymptotically flat.  The dual transformed string
coupling constant is given by
\begin{equation}
 g^{\prime}_s = g_s \frac{R_7^{\prime} R_8^{\prime} R_9^{\prime}}
  {l_s^3} = \frac{l_p^3}{R_7 R_8 R_9} ,
\label{3coup}
\end{equation}
where the dual transformed circle size $R_7^{\prime} 
= l_p^3 / (R_7 R )$ (and similarly for $R_8^{\prime}$ and 
$R_9^{\prime}$).  The string length scale $l_s$ is invariant
under the $T_7 T_8 T_9$ transformation.  In terms of these parameters, 
the function $h$ of Eq. (\ref{t3h}) can be rewritten as
\begin{equation}
 h = \frac{2 \pi^3}{15} g^{\prime}_s l_s^4 
\sum_{n=1}^{N} \frac{Q_n}{| \vec{r} - \vec{r}_n |^4}  . 
\label{wow}
\end{equation}

The distinctive feature of $T^3$ compactification is that the 
string coupling constant $g^{\prime}_s$, Eq. (\ref{3coup}),
is independent of the size of the original eleventh circle $R$ due 
to the self-duality under $S$ dual transformation.  This implies
that the effective eleventh circle size relative to the eleventh 
Planck length remains fixed even under the decompactification 
limit $R \rightarrow \infty$ of the original system.  Therefore it
is sensible to consider the ten dimensional background geometry
even in the limit $R \rightarrow \infty$.  Thus, when considering
the decompactification limit, we require the maximum value of
$Q_n$'s to asymptotically grow accordingly and set $Q = Q_{n_{max}}$
to get the function
\begin{equation}
 h =  \frac{2 \pi^3}{15} g^{\prime}_s l_s^4 \frac{Q}{r^4}
\label{wow1}
\end{equation}
where we define $r^2 = | \vec{r} - \vec{r}_{n_{max}} |^2$.
We observe that
\[ g^{\prime}_s l_s^4 Q = \left( \frac{g^{\prime 2}_s l_s^8}
   {R_7^{\prime} R_8^{\prime} R_9^{\prime} } \right)
\left( \frac{ Q R_7^{\prime} R_8^{\prime} R_9^{\prime} }
       {g_s^{\prime} l_s^4} \right) , \]
where the first factor in the curly bracket is proportional
to the seven dimensional gravitational constant and
the second factor is $Q / R$ in terms of the original
DLCQ $M$ theory parameters.  The second factor corresponds
to the energy of $Q$-times wrapped $D$ threebranes when 
the usual factor one is added to the function $h$. 
Plugging Eq. (\ref{wow1}) into Eq. (\ref{d3metric}) yields
\begin{equation}
ds_{10}^2 =  \left( \frac{2 \pi^3}{15} g^{\prime}_s  Q 
  \right)^{-1/2} \frac{r^2}{l_s^2} (-d \tau^2 +  dx_7^2 + dx_8^2 
+ dx_9^2 ) + \left( \frac{2 \pi^3}{15} g^{\prime}_s Q
 \right)^{1/2} l_s^2 ( \frac{dr^2}{r^2} + d \Omega_{(5)}^2 )
\label{ads5}
\end{equation}
where $d \Omega_{(5)}^2 $ is the metric on the unit five-sphere.
The background geometry is $AdS_5 \times S^5$ as shown in
Eq. (\ref{ads5}).

Going to $\tilde{M}$ theory side, we find that the dual 
transformed string coupling, the seventh, eighth, and
ninth dual transformed circle sizes of $\tilde{M}$ theory
are the same of those of DLCQ $M$ theory.  They accordingly
remain fixed when we take the limit 
$\tilde{R} \rightarrow 0$.  On the other hand,
the characteristic transversal length scales $\tilde{R}_i$ 
($i =1, \cdots , 6$) are proportional to $\tilde{R}^{1/2}$ and 
vanish in the same limit.  Likewise, the string length scale 
of $\tilde{M}$ theory vanishes, proportional to 
$\tilde{R}^{1/4}$ when we take $\tilde{R} \rightarrow 0$.  
Thus, the microscopic description of the DLCQ $M$ theory on $T^3$ 
is the supersymmetric Yang-Mills theory of $D$ threebranes.
This is the $CFT_4$ side description in the holographic
AdS$_5$/CFT$_4$ correspondence, and we showed that the
background geometry of DLCQ $M$ theory on $T^3$ is 
$AdS_5 \times S^5$.  

\subsection{$T^4$ : AdS$_7$/CFT$_6$ Correspondence}

Our starting point here is the limit where $R$ and $R_i$
($i=6,7,8,9$) are small.  Upon the dimensional reduction along
the eleventh light-cone circle and taking $T_6 T_7 T_8 T_9$
duality transformation, the IIA supergravity fields 
look like
\begin{equation}
ds_{10}^2 = \frac{1}{\sqrt{h}} (- d \tau^2 + dx_6^2 + \cdots
+ dx_9^2 ) + \sqrt{h} ( dx_1^2 + \cdots + dx_5^2  )
\label{d4metric}
\end{equation}
\[  e^{-2 \phi} = \left( \frac{l_p^{9/2}}{R^{1/2} R_6 R_7 R_8 R_9} 
   \right)^{-2} h^{1/2} , \]
where the five-dimensional harmonic function $h$ is
given by 
\begin{equation}
 h = \frac{\pi^4}{15} \frac{l_p^9}{R^2 R_6 R_7 R_8 R_9 }
 \sum_{n=1}^{N} \frac{Q_n}{|\vec{r} - \vec{r}_n |^3 } .
\end{equation}
Here $\vec{r}$ represents a five dimensional vector
for five non-compact dimensions $(x_1 , \cdots , x_5 )$.
The only non-vanishing RR field is the magnetic three-form
gauge field whose charge is $D$ fourbranes wrapped along
$(6,7,8,9)$ four-torus.  

We note that under the dual transformation, the dual string
coupling constant and the dual sizes of the four-torus are
given by
\begin{equation}
g^{\prime}_s = g_s \frac{l_s^4}{R_6 R_7 R_8 R_9} 
 = \frac{l_p^9}{R^{1/2} R_6 R_7 R_8 R_9} \ , \
 R^{\prime}_i = \frac{l_s^2}{R_i} = \frac{l_p^3}
 {R_i R }
\end{equation}
in terms of the original IIA parameters and $M$ theory
parameters, respectively.  The string length scale $l_s$ 
remains fixed.  We observe that, for the small values of original
light-cone $M$ theory circle size $R$ and $R_i$, the dual
string coupling is large; we should thus lift the 
solution into the eleven dimensions.  The corresponding eleven 
dimensional metric reads
\begin{equation}
ds_{11}^2 = h^{-1/3} (- d \tau^2 + dx_6^2 + \cdots
+ dx_9^2 + dx_{11}^2 ) + h^{2/3} ( dx_1^2 + \cdots + dx_5^2  )
\label{m5metric}
\end{equation}
where we rewrite the five-dimensional harmonic function $h$ as
\begin{equation}
 h = \frac{\pi^4}{15} l_p^{\prime 3}
 \sum_{n=1}^{N} \frac{Q_n}{|\vec{r} - \vec{r}_n |^3 } ,
\end{equation}
using dual transformed eleven dimensional Planck scale
\[ l_p^{\prime} = \frac{l_p^3}{R^{2/3} (R_6 R_7 R_8 R_9 
   )^{1/3} } . \]
The eleven dimensional circle that is now spatial has 
a size $R_{11}^{\prime}$ 
\[ R_{11}^{\prime} = \frac{l_p^6}{R R_6 R_7 R_8 R_9 } . \]
Eq. (\ref{m5metric}) is identical to the near horizon
geometry of the longitudinal $M$ fivebranes, which is an
exact solution of the eleven dimensional supergravity.
We note that
\[ l_p^{\prime 3} Q_n = \left( \frac{l_p^{\prime 9 }}
               {R_6^{\prime} R_7^{\prime} R_8^{\prime} 
                R_9^{\prime} R_{11}^{\prime} } \right)
\left( \frac{ Q_n  R_6^{\prime} R_7^{\prime} R_8^{\prime} 
                R_9^{\prime} R_{11}^{\prime} }{l_p^{\prime 6}} 
\right) , \]
where the first factor in the curly bracket is proportional
to the six dimensional gravitational constant and
the second factor is $Q_n / R$ in terms of the original
DLCQ $M$ theory parameters.  The second factor corresponds
to the energy of $Q_n$-times wrapped $M$ fivebranes when 
the usual factor one is added to the function $h$.  Here
it means that we are in the $Q ( \equiv Q_{n_{max}} )$-sector 
DLCQ $M$ theory in the decompactified limit $R \rightarrow
\infty$.  In this limit, we introduce $r^2 = 
|\vec{r} - \vec{r}_{n_{max}} |^2 $ and rewrite the function
$h$ as
\begin{equation}
 h = \frac{\pi^4}{15} l_p^{\prime 3} \frac{Q}{r^3} .
\label{puha}
\end{equation}
Introducing a new spatial coordinate 
$ U^2 =  r / l_p^{\prime 3} $
and plugging Eq. (\ref{puha}) into Eq. (\ref{m5metric})
yield
\begin{equation}
ds_{11}^2 =  \left( \frac{\pi^4}{15} Q \right)^{-1/3} l_p^{\prime 2} 
  U^2 (-d \tau^2 +  dx_6^2 + \cdots + dx_9^2 + dx_{11}^2 ) + 
   \left( \frac{\pi^4}{15}  Q \right)^{2/3} l_p^{\prime 2 } 
( \frac{4 dU^2}{U^2} + d \Omega_{(4)}^2 )
\label{ads7}
\end{equation}  
where $d \Omega_{(4)}^2$ is the metric on the unit four-sphere
and we use spherical coordinates.  The eleven dimensional
background geometry shown in Eq. (\ref{ads7}) is $AdS_7 \times
S^4$.

Microscopic degrees of freedom in this case can be understood
by resorting to the $\tilde{M}$ theory \cite{roz} \cite{brs}
\cite{t4micro}.  We find that the
string coupling constant and the string length scale of 
$\tilde{M}$ theory are
\begin{equation}
 \tilde{g}_s^{\prime} = \left( \frac{R}{\tilde{R}} \right)^{1/4}
  g_s^{\prime} \ , \
 \tilde{l}_s = \left( \frac{ \tilde{R}}{R} \right)^{1/4} 
 l_s .
\end{equation}
In the limit $\tilde{R} \rightarrow 0$, the string coupling
constant diverges and, as a result, we have to go to the eleven 
dimensional $\tilde{M}$ theory.  In fact, the size of the
eleventh spatial circle and the size of the dual torus
radii of $\tilde{M}$ theory are finite in the $\tilde{R}
\rightarrow 0$ limit.
\begin{equation}
 \tilde{R}_{11}^{\prime} = g_s^{\prime} l_s = 
  R_{11}^{\prime} \ , \
 \tilde{R}_i^{\prime} = R_i^{\prime}
\end{equation}
On the other hand, the characteristic transversal length
scale ($j = 1, \cdots , 5$) and the eleven dimensional 
Planck length vanish in the same limit.
\begin{equation}
 \tilde{R}_j^{\prime} = \left( \frac{\tilde{R}}{R}
       \right)^{1/2} R_j^{\prime}   \ , \
 \tilde{l}_p^{\prime} = \left( \frac{\tilde{R}}{R}
       \right)^{1/6} l_p^{\prime}
\end{equation}
The vanishing of the eleven dimensional Planck length
implies that the bulk dynamics in eleven dimensions
freezes, and we have the microscopic description on the 
world-volume of $\tilde{M}$ fivebranes, (2,0) field 
theory \cite{twozero}.  This theory has new degrees of 
freedom; the Yang-Mills coupling 
constant $g_{YM}^2 = \tilde{g}^{\prime}_s \tilde{l}_s = 
 g_s^{\prime} l_s $
is finite (when taking $\tilde{R} \rightarrow 0$ limit)
and the new degrees of freedom appear at about the energy scale 
$g_{YM}^{-2} = (g_s^{\prime} l_s )^{-1}$.  Since this is 
the inverse of the eleven dimensional circle size 
$\tilde{R}_{11}^{\prime}$, the extra degrees of freedom
corresponds to the $\tilde{M}$ momentum modes along the
$\tilde{M}$ theory circle.

These modes, in the decompactified limit where the size
of the eleventh circle becomes infinite, become massless
and they can thus contribute to the background metric
of DLCQ $M$ theory, recalling that 
$\tilde{R}_{11}^{\prime} = R_{11}^{\prime}$.  
In the type IIA limit of the $\tilde{M}$ theory, these
are contributions from $D$-particles.  Therefore, since
we have supersymmetric bound states of $D$ fourbranes and
$D$-particles, we write the type IIA DLCQ supergravity
fields as
\begin{equation}
 ds_{10}^2 = - h_0^{-1/2} h^{-1/2} d \tau^2 
 + h_0^{1/2} h^{-1/2} (dx_6^2 + \cdots + dx_9^2 )
 + h_0^{1/2} h^{1/2} (dx_1^2 + \cdots + dx_5^2 )
\label{lol}
\end{equation}
\[ e^{-2 \phi} = g_s^{\prime -2} h^{1/2} h_0^{3/2} 
  \ , \ A_{\tau} = - \frac{1}{h_0} , \]
which is an exact BPS solution of DLCQ IIA supergravity. 
Now RR one-form gauge field is non-vanishing, as well 
as magnetic RR three-form gauge field.  The function $h$ is the
same as before, and $h_0$ is given by
\begin{equation}
 h_0 = \frac{l_p^{\prime 9}}{R_6^{\prime} R_7^{\prime}
R_8^{\prime} R_9^{\prime} R_{11}^{\prime}} 
  \left( \frac{Q_0}{R_{11}^{\prime}}
 \right) \frac{1}{r^3} = \frac{Q_0 l_p^3}{r^3} 
\end{equation}
and it is another harmonic function defined in the non-compact
five dimensional space $(x_1 , \cdots , x_5)$.  
Lifting the ten-dimensional fields Eqs. (\ref{lol}) into
the eleven dimensions and taking the decompactified limit
$R_{11}^{\prime} \rightarrow \infty$ produce the following
eleven dimensional metric ($x^+ = 2 \tau$) 
\begin{equation}
ds_{11}^2 = h^{-1/3} ( dx^+ dx^- + h_0 dx^- dx^- 
 + dx_6^2 + \cdots + dx_9^2 ) + h^{2/3} 
 (dx_1^2 + \cdots + dx_5^2 ) ,
\label{modif}
\end{equation}
where the function $h_0$ has been changed into 
\begin{equation}
 h_0 \rightarrow h_0 = \frac{l_p^{\prime 9}}
{R_6^{\prime} R_7^{\prime} R_8^{\prime} R_9^{\prime} } 
\left( \frac{p_- \delta (x^- - x^-_0)}{r^3}
 \right) 
\end{equation}
as in the case of $T^0$ DLCQ $M$ theory.  Here $p_- =
Q_0 / R_{11}^{\prime}$ that is kept fixed in the limit
$R_{11}^{\prime} \rightarrow \infty$, and $x^-_0$ is 
a real parameter.  We notice that outside the support
of the delta function Eq. (\ref{modif}) is identical
to Eq. (\ref{ads7}), i.e., $AdS_7 \times S^4$, except
that the eleventh circle is now light-like.  In fact,
as we take the limit $r \rightarrow \infty$ to go to
the boundary manifold of AdS$_7$, the modified term
in Eq. (\ref{modif}) proportional to $h_0$ asymptotically 
drops out and the asymptotic geometry
is precisely that of $AdS_7 \times S^4$; the only
difference in asymptotic geometries between Eq. (\ref{modif})
and Eq. (\ref{ads7}) is that the eleventh circle is 
light-like.  In other words, the new degrees
of freedom of (2,0) field theory do not change the
asymptotic geometry and can be regarded as the fluctuation
of exact $AdS_7 \times S^4$.  This consideration heuristically
shows the validity of AdS$_7$/CFT$_6$ correspondence.

\section{Discussions}

We found a uniform derivation of the background geometry
of DLCQ $M$ theory.  For even values of $p$, the background
geometry is eleven dimensional, while for odd values of $p$
the background geometry is ten dimensional. 
For $p > 1$, the background geometry of DLCQ $M$ theory
is the tensor product of AdS$_{d+1}$ space ($d= 3, 4, 6$) and a 
sphere.  We emphasize that our analysis makes sense even for the 
case when the transversal coordinates of the AdS$_{d+1}$ space, 
i.e., the spatial 
coordinates other than the radial coordinate, form a finite
torus $T^{\prime d-1}$ in the finite $N$-sector DLCQ $M$ 
theory\footnote{In the case of single-center solutions
with the finite $N$ and finite dual torus size, the 
background geometry is
an AdS space with identifications.  In the case of the
multi-center solutions, the space-time geometry is 
asymptotically an AdS space.}. 
In this case, the number of geometric killing spinor is 
sixteen \cite{pope}, and the AdS$_{d+1}$ space has the 
diffeomorphism symmetry group that is a proper subgroup of  
$SO(d,2)$.  The microscopic
description is given by the supersymmetric Yang-Mills matrix 
theory or $(2,0)$ matrix theory.  In the limit of DLCQ $M$ theory 
where the $M$ theory 
circle is decompactified (large $N$ sector) along with 
the dual torus of $T^p$, the diffeomorphism symmetry
enhances to $SO(d,2)$ and the killing spinor number gets
doubled to 32 \cite{khl}.  On the boundary manifold of the 
$AdS_{d+1}$, which is $d$-dimensional (that is isomorphic
to the world-volume
of $(d-1)$-brane of $\tilde{M}$ theory), the preservation of 
the bulk diffeomorphism symmetry $SO(d,2)$ translates to the 
conformal symmetry in $d$ dimensions (the world-volume conformal 
symmetry of $\tilde{M}$ theory) \cite{witten}. Thus, 
in this limit, the conformal symmetric sector of the matrix theory 
side description is relevant.  Therefore, for the case of
AdS$_{4,5,7}$, we understand why AdS/CFT correspondence
works.  It is simply the correspondence between the DLCQ
$M$ theory in the decompactified limit and the $\tilde{M}$
theory in the conformal limit. 

Recently, Vafa gave explicit examples in the context of 
AdS$_3$/CFT$_2$ that show the incompleteness of AdS/CFT 
correspondence \cite{vafa}.  Our analysis in this paper do not
yield $AdS_3$ background geometry and it suggests that
the similar examples in AdS$_{4,5,7}$ should be
absent. The typical situation where $AdS_3$ (or $AdS_2 \times S^1$) 
can appear is the case of the $S$ dual version of $D$-string 
$D$ fivebrane bound states \cite{hyun} \cite{hsk}.    
In the language of $\tilde{M}$ theory on 
$T^5$ \cite{brs} \cite{ns}, this 
situation contains new degrees of freedom corresponding to 
string-like excitations in addition to the usual Yang-Mills
description \cite{seiberg}.  It remains to be seen 
if this extra complication leads to the explanation of 
Vafa's puzzle.  We note that for the description of
these `micro-strings' \cite{dvv2}, we possibly need
an additional infinite boost that should realize
the nine-eleven flip for these micro-strings, and the simple 
picture of holography given in this paper should be modified. 

For $p = 0, 1$, the submanifold of the support of the 
parton momentum ($M$-momentum in BFSS theory and the fundamental 
string momentum in DVV theory) are $R^{9-p}$.  The target space 
conformal symmetry of $R^{9-p}$ manifold is $SO(10 - p , 1)$.
Recalling that the background geometry is $R^{10-p ,1}$,
the target space conformal symmetry is identical to the Lorentz
symmetry of the background geometry.  Further elucidation
of this point will help us better understand the
covariance of the matrix theory in the infinite momentum frame.

It is interesting to note that the curved geometries of AdS spaces 
preserve the full 32 supersymmetry in the decompactified limit 
and they are classically (curvature) singularity-free.  
In our opinion, these space-times provide us with true 
space-time vacua of DLCQ $M$ theory.  To sharpen this 
assertion, it is important to compare the scattering 
calculations of the matrix theory on $T^p$ ($p>1$) to the 
supergravity calculations on the AdS background.  The 
investigation in this regard will be reported 
elsewhere \cite{hsk2}.

There are several important lines of generalization of 
this paper.  The DLCQ $M$ theory on $T^6$ \cite{t6lit}
poses a challenging problem in its microscopic description 
due to the appearance of the infinite number of new degrees 
of freedom.  However, as far as the background geometry is 
concerned, we can follow the procedure of this paper 
to obtain the Kaluza-Klein monopole background in 
eleven dimensions.  It remains to be seen what this background
geometry teaches us about the microscopic description.
For the DLCQ $M$ theory on $T^p$ with $p>1$, we can consider
the case when the torus is non-commutative \cite{noncom}.
This generalization, that requires non-vanishing
NS-NS two-form gauge fields, is essential to uncover,
for example, the full structure of $(2,0)$ matrix theory
including Fayet-Iliopoulos terms.  In addition, the heterotic 
side generalization of our analysis will obviously be of 
interest.

\acknowledgements{We would like to thank Jae-Suk Park 
and Hyeonjoon Shin for helpful discussions.}

\end{document}